\documentclass[preprint,11pt]{revtex4-1}

\usepackage{amsmath,amssymb}
\usepackage{hyperref}
\usepackage{verbatim}
\usepackage{graphicx}

\renewcommand{\phi}{\varphi} 
\newcommand{\ket}[1]{\left|#1\right\rangle}

\newcommand{\ketbra}[2]{\left|#1\right\rangle \left\langle#2\right|}

\usepackage{xcolor} 
\definecolor{darkblue}{rgb}{0.0, 0.0, 0.55}
	\hypersetup{%
	pdfborder = {0 0 0},
	colorlinks = true,
	bookmarksdepth = section,
	citecolor = darkblue,
	linkcolor = darkblue,
	urlcolor = darkblue,
	}%

\begin{document} 

\title{Computing LDOS resonance energy shifts of monatomic doped chains}

\author{R. N. P. Maia}
\affiliation{Universidade Federal do Rio de Janeiro, Campus Maca\'e, 27930-560 Maca\'e, Rio de Janeiro, Brazil}

\author{C. M. Silva da Concei\c{c}\~ao}
\affiliation{Universidade Federal Fluminense, RHS/RCN, 28895-532 Rio das Ostras, Rio de Janeiro, Brazil}
  
\date{\today}

\begin{abstract}

An analytical method to compute the LDOS energy spectrum and stationary states for finite size doped monatomic chains modelled by an effective one-dimensional tight-binding hamiltonian is presented. It is based on the formal solution of linear second order recurrence relations. 
We also study the LDOS energy spectrum of some doped monatomic chains applying a perturbative approach to the characteristic equation of the reference metallic structure doped with a few extraneous atoms.

\end{abstract}

\maketitle
 
Certain monatomic metallic chains on a specific substrate displays one-dimensional (1D) electronic states which are decoupled from the bulk band structure \cite{onc08}. As reported for linear Au chain \cite{nil02}, Pd chain \cite{nil05a} or Au chain with Pd impurities \cite{wal05,nil05b} built on NiAl(110), ressonances of the local density of states (LDOS) were measured using low-temperature scanning tunneling microscopy (LT-STM). In order to describe the 1D quantum confinement of pure atomic chains, Nilius  \emph{et al} considered the energy spectrum and the eigenstates of an 1D quantum well, with a parabolic dispersion relation. The latter was fitted from the wave number, which in turn was obtained from the nodes of the LDOS ressonances.

Further experimental investigations with the LT-STM \cite{fol04,fol04b} showed that LDOS electronic ressonance states decoupled from the bulk band structure arise at low-temperature even for a homogeneous metallic assembly with the same substrate and adatoms formed by an atomic chain of Cu on Cu(111). Specifically for these homogeneous systems, F\"{o}lsch \emph{et al} reported resonance states well described by linear combinations of some proper hybridized atomic orbitals, i.e., the system LDOS can be modelled as an artificial molecule of finite size in the H\"uckel approximation. This claim was justified after they measured and fitted a single band dispersion relation that is  well described by an effective 1D tight-binding model with a finite size tridiagonal hamiltonian matrix containing two parameters: the binding energy and the hopping integrals. 
A similar behavior of resonance states was observed later for the LDOS of some monatomic Cu chain doped with a few cobalt atoms \cite{fol07}. Four parameters were used in order to derive a model with a finite size tridiagonal hamiltonian matrix: the binding energies of Cu and Co along with the  hopping integrals for Cu-Cu and Cu-Co couplings. Surprisingly, when comparing the energy levels of doped structures with the monatomic Cu reference chain, a systematic downward shift in  all resonance energies was observed.

The purpose of this work is to explore 1D tight-binding models describing the LDOS resonances in doped chains. We also investigate if changes in energy level of the reference chain lead a systematic upward, downward or mixed shifts. Specifically, we perform a series expansion in order to obtain first order correction to the doped ($d$) energy levels $E_k^{(d)} = E_k^{(0)}+\lambda_k \epsilon +\mathcal{O}(\epsilon^2)$, where $E_k^{(0)}$ is the kth energy level for the reference chain and $\epsilon$ is the perturbative parameter.

Theoretical investigation of tridiagonal hamiltonian matrices has been reported in \cite{ban13}. For nearest-neighbor monatomic chains it was shown how to extract the discrete energy spectrum from a secular equation which encompass the Chebyshev polynomials of the second kind. Energy eigenstates were obtained from recurrence equations involving these polynomials.
 
In this paper the energy spectrum and eigenstates of the 1D tight-binding model of finite size monatomic chains are written in terms of  generalized Fibonacci polynomials (GFP) \cite{sha09,amd14}. The discrete set of LDOS ressonance energies is obtained from the roots of a GFP with proper constant coefficients, while the energy eingenstates in the basis of local orbitals \cite{gro13} has coefficients given by GFP. 
To the authors knowledge there is no systematic analysis concerning linear second order recurrence relations and GFP for the description of 1D tight-binding hamiltonian. Interestingly, the same kind of recursive relations also appear for the finite-size partition function of Ising spin 1/2 systems with nearest-neighbor couplings \cite{con17}.
 
To treat a general 1D doped monatomic chain, one should solve a linear second-order recurrence relation with nonconstant coefficients in order to find the eigenstates. For that we use a general recursive approach introducing the nonhomogeneous generalized Fibonacci polynomials (NHGFP), which further extend the generalized Fibonacci polynomials.  
Upon this analysis we found that the energy spectrum can be computed from a secular equation which includes those NHGFP. 
Moreover, we apply our theoretical results to simple monatomic structures in order to compute the shift in the energy spectrum due to one or two extraneous atoms included in a reference chain. We also show that reflection symmetry of a single doped monatomic structure implies that the antisymmetric part of the energy spectrum has no ressonance energy shift. It is true when comparing a specific single cobalt doped symmetric monoatomic Cu chain with respect to the reference monatomic Cu chain. It has been experimentally probed that there is no shift within experimental precision for the first excited resonance state \cite{fol07}.


Let us consider quantum confinement in atomic chains, i.e., a quantum particle inside a 1D sample \cite{gro13} of equally spaced atoms at site $na$ with lattice parameter $a$ and local orbital $\Phi_a$ or any other quantum system described by an effective hamiltonian operator as the LDOS of monatomic chains \cite{fol04,fol04b,fol07}. The 1D nearest-neighbor tight-binding (TB) hamiltonian reads
\begin{gather}\label{ctb}
\hat{H}_N = \sum_n \left\{
\alpha_n \, \hat{\Pi}_{n,n} 
+ \gamma_{n,n+1} \, ( \hat{\Pi}_{n,n+1} + \hat{\Pi}_{n+1,n} ) \right\}
\, ,
\end{gather}  
where $\ket{n} = \ket{\Phi_a (x-na)}$ is a complete set of orthogonal states and $\hat{\Pi}_{nm} = \ketbra{n}{m}$ are projectors. The caracteristic energies are the binding energies $\alpha_{n}$ and the nearest-neighbor hopping integrals $\gamma_{n,n+1}$. It is noticeable that the TB hamiltonian defined in Eq. \eqref{ctb} is the most general tridiagonal matrix, which by the Lanczos method \cite{gro13} can represent any 1D finite size quantum system.

It is convenient to use the kth energy eingestate $\ket{E(k)}$, such that the time-independent Schr\"odinger equation is written in terms of a second order recurrence relation for the probability amplitudes $c_n (k)$
\begin{gather}\label{tbam:timeindependent}
c_{n+1} (k) = p_{n} (k) \, c_n (k) + q_{n} \, c_{n-1} (k) 
\, ,
\end{gather}
with site-dependent coefficients $p_{n} (k) = ( E(k) - \alpha_n )/\gamma_{n,n+1}$, $q_{n} = - \gamma_{n,n-1}/\gamma_{n,n+1}$ ($n<N$) and $q_N=-1$.
For the general case the main task is computing the stationary states by solving the recurrence relation \eqref{tbam:timeindependent}. In the particular case where these coefficients are site-independent one can find the energy spectrum and the corresponding eigenstates in the Wannier basis in terms of Chebyshev polynomials \cite{ban13} or, as described in this paper, in terms of generalized Fibonnaci polynomials (GFP) \cite{amd14,sha09}.


The general case with nonconstant coefficients (\ref{tbam:timeindependent}) can be solved by introducing the nonhomogeneous generalized Fibonacci polynomials (NHGFP) denoted by $F_{N+1}(P;Q)\equiv F_{N+1}(p_{1},\dots,p_{N};q_{2},\dots,q_{N})$, 
where a useful compact notation is used  $P\mapsto(p_{n})_{n=1}^{N}= (p_{1},\ldots,p_{N})$ and $Q\mapsto(q_{n})_{n=2}^{N}=(q_{2},\ldots,q_{N})$.
Similar to the GFP, these NHGFP are generated recursively through the second-order recurrence relation
\begin{gather}\label{nhgfp:rr}
F_{n+1}(P;Q)=p_{n}F_{n}(P;Q)+q_{n}F_{n-1}(P;Q)\, ,
\end{gather} 
 with initial polynomials $F_{0}(P;Q)=0$ and $F_{1}(P;Q)=1$. For instance, we can explicitly compute some low-order polynomials $F_{2}(p_{1})  =  p_{1}$, $F_{3}(p_{1}, p_{2}; q_{2}) = p_{2} p_{1} + q_{2}$, $F_{4}(p_{1}, p_{2}, p_{3}; q_{2}, q_{3}) = p_{3} p_{2} p_{1} + p_{3} q_{2} + q_{3} p_{1}$ and so on.

The formal solution for the recurrence relation \eqref{tbam:timeindependent} seems not to have been previously reported. It can be written explicitly in terms of the NHGFP and its derivatives:
\begin{equation}\label{amplirr}
c_N = \left[F_{N}(P;Q)\right] \, c_1 +\left[ q_1\, \frac{ \partial }{ \partial p_1 }\, F_{N}(P;Q)\right] \, c_0
\, .
\end{equation}
Henceforth we specify  boundary conditions considering a finite-size 1D atomic chain sample, where one must impose that $\gamma_{10} = \gamma_{N,N+1} = 0$ at the edges of the chain. These constraints necessarily provide $q_{1}=0$, $p_{N}=(E(k)-\alpha_{N})/\gamma_{N,N-1}$ and $q_{N}=-1$. 
The confinement of the particle in $N$ localized orbitals entails a discrete energy spectrum because after applying the boundary conditions $c_0 = c_{N+1} = 0$ in Eq. \eqref{amplirr} we arrive at $c_{N+1}=\left[F_{N+1}(P;Q)\right] \, c_{1} = 0$. Thus in order to find a nontrivial solution it suffices that
\begin{equation}\label{atomic-chain-bc}
F_{N+1}(P;Q) = 0  
\, ,
\end{equation}
since $c_1$ must not vanish. Note that $F_{N+1}(P;Q)$ is a polynomial of order $N$ of the energy $E_k$. 
From the secular equation \eqref{atomic-chain-bc}, the  roots of a NHGFP provides the allowed energies of the one-dimensional atomic chain.


Since a monatomic chain will serve as a reference chain for our later analysis, let us briefly consider the homogeneous TB model with equal binding energies $\alpha_n = \alpha$ and hopping integrals $\gamma_n = \gamma$ for all integer $n$. In this particular case the stationary states for the quantum particle are recursively generated by Eq. \eqref{tbam:timeindependent} whose formal solution is given by Eq. \eqref{amplirr} with $c_0 = 0$. The proper coefficients are $p_n (k) \equiv p = ( E(k) - \alpha ) / \gamma$ and $q_n = q = -1$, since both parameter sequences collapse: $P\mapsto (p,\ldots,p)\equiv p$ and $Q\mapsto (-1,\ldots,-1)\equiv -1$ 

The recurrence relation \eqref{nhgfp:rr} is exactly the one obeyed by the generalized Fibonacci and Lucas polynomials \cite{amd14,sha09}. 
The only difference here are the seeds $c_0$ and $c_1$ to generate the energy eigenstates, which in general can be complex numbers. 
In order to prove this claim it suffices to consider the generalized Fibonacci polynomials (GFP) $F_n(p,q)$ in real variables $p$ and $q$, which are recursively generated by $F_{n+1} (p,q) = p \, F_n (p,q) + q\, F_{n-1} (p,q)$,
with initial seeds $F_0 = 0$ and $F_1 = 1$. 
The recursive solution has a Binet form $F_n (p,q) = ( \Phi_+^n - \Phi_-^n )/(\Phi_+ - \Phi_- )$ with $\Phi_\pm (p,q) = ( p \pm \sqrt{ p^2 + 4 q } )/2$. 
The final form of the stationary states can be written as a linear combination of the two functions $\Phi_\pm (p,q)$, which after substitution in order to fulfill the proper seeds $c_0=0$ and $c_1$ is given by $c_n (k) = F_n (p_k,-1) \, c_1$.

A closed-form solution for the energy spectrum can be obtained if one solves the energies $E_k = E(k)$ in variable $p_k = p(k)$ of the GFP of degree $N$ throught the characteristic equation $F_{N+1}(p,-1) = 0$. These roots can be easily computed from the Binet form with the  constraint $F_{N+1}(p,-1) = 0$,  $[\Phi_+(k)/\Phi_-(k)]^{N+1} = e^{2ik\pi}$, and $\Phi_+(k) \neq \Phi_-(k)$. Afterward we can use the branches of the $N+1$ root of unit such that $\Phi_\pm (k) = e^{\pm i\phi_k}$ with $\phi_k = k\pi/(N+1)$, and then $p_k = ( E_k - \alpha ) / \gamma = 2\cos\phi_k$ is a quantized variable with quantum number $k$. Finally the energy spectrum of the reference chain takes the form: 
\begin{gather}\label{enspecX}
E_k = \alpha + 2 \gamma \, \cos \phi_k \quad  (1 \leq k \leq N)
\, .
\end{gather}
This spectrum is described elsewhere \cite{ban13}, and appears from the cosine of $N+1$ equidistant partition of the positive half-plane with angles $\pi/(N+1)$ times $2\gamma$ centered at the onsite energy $\alpha$. Moreover, the coefficients in Eq. \eqref{tbam:timeindependent} for the energy eigenstates of the reference chain are given by $c_n (k) = F_n (p_k,-1) = \sin(n\phi_k)/\sin \phi_k$.


We now analyze hereafter the energy levels of some simple doped monatomic chains. These are given by the roots of Eq. \eqref{atomic-chain-bc} as modifications of the energy levels of the reference chain given by \eqref{enspecX}. The first and second structures are monatomic X chains doped with a single extraneous Y atom, and the third one is doped with two extraneous Y atoms. These chains exhausts all kinds of structures investigated experimentally in \cite{fol07}. We denote the binding energies for X atoms by $\alpha_x$ and for $Y$ atoms by $\alpha_y$, and hopping integrals $\gamma_{xx}$ and $\gamma_{xy}$. There is no hopping integral $\gamma_{yy}$ in our analysis since we do not consider any chain with two neighboring Y atoms. For all plots in this paper we use data from \cite{fol04,fol04b,fol07}, where $\alpha_x = 3.31 \text{ eV}$, $\alpha_y - \alpha_x = \epsilon = -0.35 \text{ eV}$, $\gamma_{xx} = -0.95\text{ eV}$ and $\gamma_{xy} = -0.94\text{ eV}$. It is worth noting that the hopping integrals are nearly equals, with energies ratio of 1\%.

\begin{figure}[htbp!]
\begin{center}
\includegraphics[width=0.6\columnwidth]{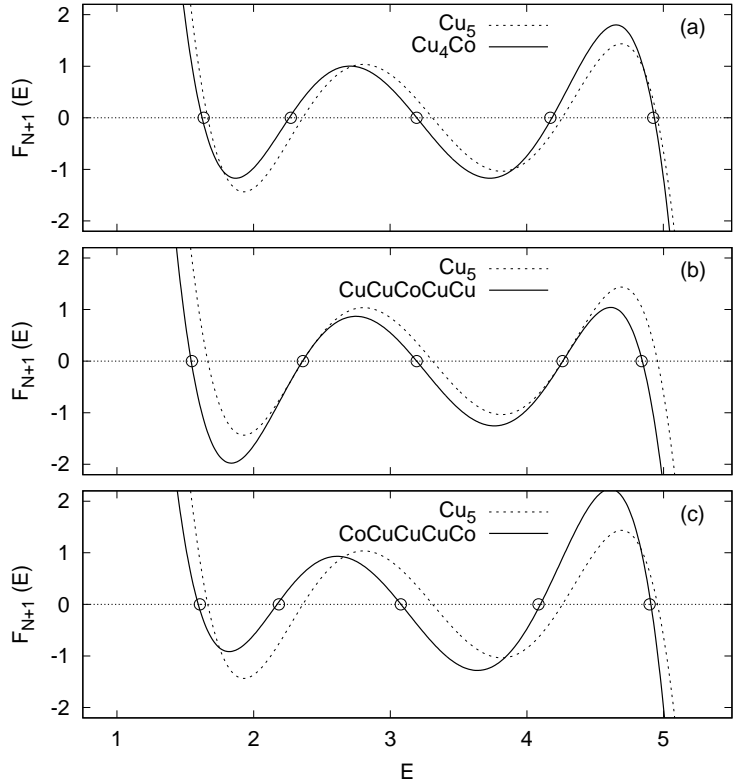}
\caption{\label{xyfig} Polynomials of order $N+1$ for monatomic chain X$_{N}$ (gray lines) and doped chain (black lines) with size $N=5$: (a) X$_{N-1}$Y; (b) X$_\frac{(N+1)}{2}$YX$_\frac{(N+1)}{2}$; (c) YX$_{N-2}$Y. First order corrections for the energy levels of the monatomic chain determined by Eqs. \eqref{enspecX}, \eqref{lambda2} and \eqref{yxy:lambda} are marked by gray circles at the zero axis.} 
\end{center}  
\end{figure}

The first example is a chain of type X$_{N-1}$Y of  size $N$, using the dimensionless parameters $p_n = p_x$ $(n<N)$, $p_N = p_y$, $q_n = -1$ $(n \neq N-1)$ and $q_{N-1} = - \gamma_{xx}/\gamma_{xy}$. It should be noted that the analysis for this structure will be the same for the mirrored chain YX$_{N-1}$. 
Figure \ref{xyfig} (a) shows the two polynomials GFP (gray lines) and NHGFP (black lines) of degree $N+1$ for size $N=5$, and the zero-axis dashed line where one can clearly observe the roots of both structures. The roots of each of these polynomials are the energy spectrum of the corresponding system. 
After performing a numerical analysis using data from \cite{fol04,fol04b,fol07} we observe that (i) the doping causes a nonequal downward shifts of all roots; (ii) it is noticiable that the energies near the edges of the spectrum shift less than  near the center of the spectrum; (iii) increasing the size of the system makes the doping effect less dicernible as one would expect.

The second single doped monatomic chain is the symmetric structure X$_\frac{N-1}{2}$Y X$_\frac{N-1}{2}$ of odd size $N$ and with the Y atom at $n=(N+1)/2$. We use the following parameters: $p_n = p_x$ for $n\neq (N+1)/2$, $p_{(N+1)/2} = p_y$; $q_n = -1$ for $n \neq (N-1)/2,(N+3)/2$, $q_{(N-1)/2} = - \gamma_{xx}/\gamma_{xy}$ and $q_{(N+3)/2} = - \gamma_{xy} / \gamma_{xx}$. Due to the symmetry of the hamiltonian for this structure, the energy spectrum splits into two subsets: the symmetric and antisymmetric LDOS eigenstates. Therefore it is expected that some antisymmetric states exhibit a node at the extraneous Y atom location, which will not differ from the respective state of the reference structure, i.e., we shall notice a vanishing resonance energy shift as is apparent from experimental observations \cite{fol07}. 
For illustration we show in Figure \ref{xyfig} (b) the plot for monatomic chain X$_\frac{N-1}{2}$Y X$_\frac{N-1}{2}$ with size $N=5$ of both polynomials GFP (gray lines) and NHGFP (black lines) of order $N+1$, and the zero-axis an a dashed line. One can clearly see the downward shift of almost roots ($k=1, 3, 5$). As expected that the energies of the antisymmetric eigenstates are unaltered ($k=2,4$).  

Finally, we consider a twice doped symmetric chain of the type YX$_{N-2}$Y of size $N$. For this structure the dimensionless parameters are $p_n = p_x$ $(n\neq 1,N)$, $p_{1} = p_{N} = p_y$, $q_n = -1$ $(n \neq 2,N-1)$, $q_2 = - \gamma_{xy}/\gamma_{xx}$ and $q_{N-1} = - \gamma_{xx}/\gamma_{xy}$. 
Figure \ref{xyfig} (c) shows both polynomials GFP (gray lines) and NHGFP (black lines) of order $N+1$ of size $N=5$, and the zero-axis dashed line where the downward shift of all roots is obvious. The same overall effect in Fig. \ref{xyfig} (a) is observed in Fig. \ref{xyfig} (c), where the energies in the center of the energy spectrum are more modified than those energies near the edge of the spectrum. However, these curves (black and gray lines) are more far away than the curves in Fig. \ref{xyfig} (a) because two extraneous Y atoms provides greater perturbation. 


We now investigate the effect of adding a few extraneous atoms in a monatomic chain with a small difference in the binding energies and equal hopping integrals. 
In general, the energy spectrum of a single doped monatomic chain system can be extracted from a secular equation which encompasses a particular NHGFP
\begin{gather}\label{sd:NHGFP}
F_{N+1} (p,\epsilon) \equiv F_{N+1} (\{p, \ldots , p , p_m , p , \ldots p\}; -1) = 0
\, ,
\end{gather}
where the extraneous Y atom occupies the site $m$ on the chain $(1 \leq m \leq N)$. This system has binding energy $\alpha_x$ everywhere except at $m$ where $\alpha_m = \alpha_y$, and, for simplicity, we assume $\gamma \equiv \gamma_{xx} \approx \gamma_{xy}$. Hence the dimensionless parameters are $p_n = p$ ($n\neq m$), $p_m = p_y$ and $q=-1$ ($1 \leq n\leq N$). The binding energy shift $\epsilon = (\alpha_y - \alpha_x)/\gamma$ between atoms X and Y yields the relation: $p_m = p - \epsilon$. The recurrence equation \eqref{nhgfp:rr} can be used in order to express the secular equation \eqref{sd:NHGFP} in terms of only three generalized Fibonacci polynomials. In fact: $F_{m+1}(\{p,\ldots,p,p_m\};-1) = p_m \, F_m (p,-1) - F_{m-1} (p,-1)$. Substituting $p_m = p - \epsilon$, this polynomial is given by $F_{m+1}(\{p,\ldots,p,p_m\};-1) = F_{m+1} (p,-1) - \epsilon \, F_{m} (p,-1)$. 
Performing the remaining $N+1-m$ iterations we arrive at $F_{N+1} (p,\epsilon) = F_{N+1} (p,q) - \epsilon \, F_{N+1-m} (p,-1) \, F_m (p,-1)$. 
Thereby a power series expansion in dimensionless parameter $\epsilon$ for the kth root $p_k$ of the polynomial \eqref{sd:NHGFP} around $p_k^{(0)} = 2\cos(\phi_k)$ can be employed straightfoward: $p_k = p_k^{(0)} + \lambda_k \, \epsilon + \mathcal{O}(\epsilon^2)$. The first order correction for the roots is controlled by the parameter \cite{wil84}:
\begin{gather}\label{lambda}
\lambda_k^\text{(sd)} = \frac{ dp_k }{ d\epsilon }
= \frac{ F_{N+1-m} (p;-1) \, F_{m} (p;-1) }{ F_{N+1}' (p;-1) } \Big|_{p=p_k^{(0)}}
\, .
\end{gather}
After substitution of $F_n (p;-1) = \sin(m\phi_k)/\sin(\phi_k)$ in Eq. \eqref{lambda} we find the following result:
\begin{gather}\label{lambda2}
\lambda_k^\text{(sd)} = \frac{2}{N+1} \, [ \, \sin(m\phi_k) \, ]^2 \geq 0
\, .
\end{gather}
A spectrum is structurally stable only if $\lambda_k \epsilon$ is definite. This is exactly the case of $\lambda_k^\text{(sd)}$ in Eq. \eqref{lambda2}, and hence $\lambda_k^\text{(sd)} \epsilon \ll 1$. Moreover, $\lambda_k^\text{(sd)} \geq 0$, thus resonance energy shifts accordingly with the sign of the perturbative parameter $\epsilon$.

\begin{figure}[htbp!]
\begin{center}
\includegraphics[width=0.6\columnwidth]{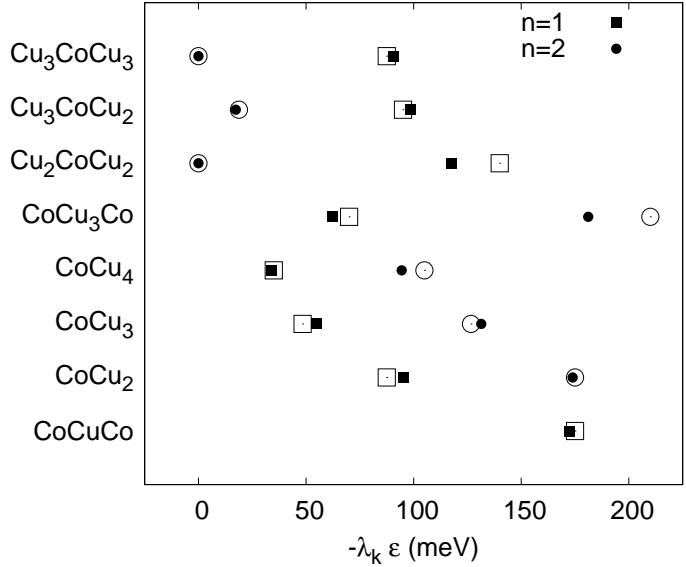}
\caption{\label{lambfig} 
First order corrections for the ressonance energy shifts $\lambda_k \epsilon$ (open symbols) given by Eqs. \eqref{lambda2} and \eqref{yxy:lambda} for various Co-Cu chains compared to the exact diagonalization using experimental data given in \cite{fol07} (black symbols).}
\end{center}
\end{figure}

Let us apply Eq. \eqref{lambda2} for the structure X$_{N-1}$Y shown in Fig. \ref{xyfig} (a), the secular equation \eqref{sd:NHGFP} 
simplifies into $F_{N+1} (p;-1) - \epsilon F_{N} (p;-1) = 0$. The first order correction for the roots is controlled by the dimensionless parameter $\lambda_k^\text{(sd1)} = 
2 \, (\sin \phi_k)^2 /(N+1)$. 
More interestingly, it can be seen that $\lambda_k^\text{(sd1)}$ is greater near the center of the spectrum where $k \sim N/2$, while it is smaller near the edge of the energy spectrum. 
This energy spectrum is shown in Figs. \ref{xyfig} (a) as gray circles in the zero axis, and in Fig. \ref{lambfig} for the ground states and first excited states for various Co-Cu chains.

A distinct behavior of the LDOS energy shifts provided by the symmetric structure X$_{(N+1)/2}$YX$_{(N+1)/2}$ shown in Fig. \ref{xyfig} (b). 
For these structure the characteristic equation \eqref{sd:NHGFP} becomes $F_{N+1} (p;-1) - \epsilon \left[ F_{\frac{N+1}{2}} (p;-1) \right]^2 = 0$. Hence the first order correction for the energies of this doped structure is given by $
\lambda_k^\text{(sd2)} =  2 \, \sin^2(k\pi/2) / (N+1)$.
A nontrivial behavior is found since there are unaltered roots, i.e., vanishing sensibility parameter $\lambda_k = 0$ for $k=2l$ $(l=1,2,\ldots, (N-1)/2)$ and non-vanishing $\lambda_k = 2/(N+1)$ for $k=2l+1$ $(l=0,1,\ldots, (N-1)/2)$.  
The first order corrections to the kth energy is shown in Fig. \ref{xyfig} (b) as gray circles in the zero axis, and in Fig. \ref{lambfig} for the two first energies of some single doped Cu chains. 
We remark that vanishing first order correction does not guarantee that high order corretions will also vanish. Moreover, it can be verified that the antisymmetric eigenvalues ($k$ even) in Eq. \eqref{enspecX} are roots of the characteristic equation, which entails that there is no ressoance shift for these specific energies.

The twice doped (td) monatomic chain YX$_{N-2}$Y shown in Fig. \ref{xyfig} (c) can be analysed interactively since the secular equation $F_{N+1} (\{ p-\epsilon , p , \ldots , p , p-\epsilon \};-1) = 0$ simplifies to $F_{N+1} (p;-1) - 2\epsilon \, F_{N} (p;-1)+\mathcal{O}(\epsilon^{2}) = 0$. The approximate solution is given by
\begin{gather}\label{yxy:lambda}
\lambda_k^\text{(td)} = \frac{4}{N+1} \, (\sin \phi_k )^2
\, .
\end{gather}
This result is twice the control parameter found for structures X$_{N-1}$Y. It is shown in Figs. \ref{xyfig} (c) as gray circles in the zero axis, and in Fig. \ref{lambfig} for low energies of various Co-Cu chains. One can conclude straightforward that $\lambda_k \epsilon \ll 1$, and so the energy spectrum of this perturbed chain is structurally stable. Since $\lambda_k^\text{(td)}$ is a positive quantity, the LDOS energy shifts follow the sign of the perturbative parameter $\epsilon$.


In summary, the method presented in this paper allows to investigate the energy spectrum of monatomic reference chains through the roots of GFP and of doped atomic chains looking for the roots of NHGFP. It clarifies that for the studied doped monatomic structures the energy shifts follows the sign of the perturbative parameter $\epsilon$ given by the binding energy difference. In the particular case of a monatomic Cu chain doped with Co atom(s) studied in \cite{fol07} we find a downward LDOS energy shifts and a structurally stable energy spectrum. In order to account for the effect of unequal hopping integrals, it was verified throught the canonical time-independent perturbation theory that the energy spectrum is structurally stable and follows the sign of a modified perturbative parameter.  Finally, our method could be applied in other circumstances, and thus future experimental research of one-dimensional atomic chains on surfaces can be guided in order to find the LDOS resonances.




\end{document}